\documentclass[preprint,showpacs,preprintnumbers,amsmath,amssymb,superscriptaddress]{revtex4}

\include{ams}
\usepackage{amsmath,amssymb,amsthm}
\usepackage{dcolumn}
\usepackage{bm}
\usepackage{graphicx}

\begin{document}

\title{Effect of symmetry on the electronic structure of spheroidal
fullerenes in a weak uniform magnetic field}

%
%

\author{M. Pudlak}
\email{pudlak@saske.sk}
\affiliation{Institute of Experimental Physics, Slovak Academy of Sciences, Watsonova 47,043 53 Kosice, Slovak Republic}

\author{R. Pincak}
\email{pincak@saske.sk}
\altaffiliation[on leave from ]{Institute of Experimental Physics, Slovak Academy of Sciences, Watsonova 47,043 53 Kosice, Slovak Republic}
\affiliation{Joint Institute for Nuclear Research, BLTP, 141980 Dubna, Moscow region, Russia}

\author{V.A. Osipov}
\email{osipov@thsun1.jinr.ru}
\affiliation{Joint Institute for Nuclear Research, BLTP, 141980 Dubna, Moscow region, Russia}

\pacs{36.40.Cg, 33.55.Be, 71.20.Tx}

\date{\today}

\begin{abstract}
The effect of a weak uniform magnetic field on the electronic
structure of slightly deformed fullerene molecules is studied
within the continuum field-theory model. It is shown that fine
structure of the electronic energy spectrum is very sensitive to
the orientation of the magnetic field. In particular, we found
that the magnetic field pointed in the x direction does not
influence the first electronic level whereas it causes a splitting
of the second energy level. This behavior differs markedly from
the case of the magnetic field pointed in the z direction.

\end{abstract}

\maketitle

\section{Introduction}

Recently, we have considered the problem of the low energy
electronic states in spheroidal fullerenes~\cite{Pudlak1} as well
as the influence of a weak uniform external magnetic field pointed
in the z direction~\cite{Pudlak2}. The main findings were a
discovery of fine structure with a specific shift of the
electronic levels upwards due to spheroidal deformation and
the Zeeman splitting of electronic levels due to a weak
uniform magnetic field. In addition, it was shown that the
external magnetic field modifies the density of electronic states
and does not change the number of zero modes.
In this paper, we examine the case of the magnetic field pointed in
the x direction. It is interesting to note that the obtained modification
of the electronic spectrum of the spheroidal fullerenes differs
markedly from the case of the z-directed magnetic field. The reason is the
proper symmetry of the spheroid, which changes the role of the external
magnetic field in comparison with the spherical case.
This gives an additional possibility for experimental study of
the electronic structure of deformed fullerene molecules.

Notice that the Schr\"odinger equation for a free electron on the
surface of a sphere in a uniform magnetic field was formulated and
solved to describe Zeeman splitting and Landau quantization of
electrons on a sphere in Ref.~\cite{Aoki}. We have explored in the
Ref.~\cite{Pudlak2} the field-theory model where the specific
structure of carbon lattice, geometry, and the topological defects
(pentagons) were taken into account. Following the Euler's theorem
one has to insert twelve pentagons into hexagonal network in order
to form the closed molecules. In the framework of continuum
description we extend the Dirac operator by introducing the Dirac
monopole field inside the spheroid to simulate the elastic
vortices due to twelve pentagonal defects. The exchange of two
different Dirac spinors which describes the K spin flux in the
presence of a conical singularity is included in a form of
t'Hooft-Polyakov monopole. Our studies cover slightly elliptically
deformed molecules in a weak uniform external magnetic field
pointed in the x directions.

\section{The model and the results}

Spheroidal fullerenes can be considered as an initially flat
hexagonal network which has been wrapped into closed monosurface by using
of twelve disclinations~\cite{Kroto}. We start from the tight-binding
model of graphite layer with the trial wave function taken in the form
\begin{equation}
\chi(\vec{r})=\psi_{A}(\vec{r})\chi_{A}(\vec{K},\vec{r})+\psi_{B}(\vec{r})\chi_{B}(\vec{K},\vec{r}).
\end{equation}
As is seen, the trial function is described by smoothly varying
envelope functions $ \psi_{A,B}(\vec{r})$ multiplying by the Bloch
functions $\chi_{A,B}(\vec{K},\vec{r})$. Within the $\vec{k}\vec{p}$
approximation one obtains the equations algebraically identical to a
two-dimensional Dirac equations, where two component wave function
$\psi $ represents graphite sublattices $A$ and $B$. Following the
approach developed in~\cite{Gonzales,Osipov} let us write down the
Dirac operator for free massless fermions on the Riemannian spheroid
$S^{2}$. The Dirac equation on a surface $\Sigma$ in the presence of
the abelian magnetic monopole field $W_{\mu}$ and the external
magnetic field $A_{\mu}$ is written as~\cite{Davies}
\begin{equation}
i\gamma^{\alpha}e_{\alpha}^{\ \mu}[\nabla_{\mu} - iW_{\mu}-
iA_{\mu}]\psi = E\psi, \label{eq:1}
\end{equation}
where $e_{\alpha}^{\mu}$ is the zweibein,
$g_{\mu\nu} = e^{\alpha}_{\mu}e^{\beta}_{\nu} \delta_{\alpha \beta}$ is the metric,
the orthonormal frame indices $\alpha,\beta=\{1,2\}$, the coordinate indices
$\mu,\nu=\{1,2\}$, and $\nabla_{\mu}=\partial_{\mu}+\Omega_{\mu}$ with
\begin{equation}
\Omega_{\mu}=\frac{1}{8}\omega^{\alpha\ \beta}_{\ \mu}
[\gamma_{\alpha},\gamma_{\beta}] \label{eq:2},
\end{equation}
being the spin connection term in the spinor representation (see
\cite{Nakahara,Gockeler} for details). The energy in (\ref{eq:1}) is
measured from the Fermi level.

The model (\ref{eq:2}) allows us to study the structure of electron levels
near the Fermi energy. It is convenient to consider this problem by using
of the Cartesian coordinates $x$, $y$, $z$ in the form
\begin{equation}
x=a~\sin\theta \cos\phi; \quad y=a~\sin\theta \sin\phi; \quad
z=c~\cos\theta. \label{eq:16}
\end{equation}
The Riemannian connection reads (cf. (4))
\begin{equation}
\omega{_{\phi 2}^{1}}=-\omega{_{\phi 1}^{2}}=\frac{a
\cos\theta}{\sqrt{a^{2}\cos^{2}\theta+c^{2}\sin^{2}\theta}}\ ; \quad
\omega{_{\theta 2}^{1}}=\omega{_{\theta 1}^{2}}=0. \label{eq:17}
\end{equation}
Within the framework of the perturbation scheme the spin connection coefficients
are written as
\begin{equation}
\omega{_{\phi 2}^{1}}=-\omega{_{\phi
1}^{2}}\approx\cos\theta(1-\delta\sin^{2}\theta), \label{eq:18}
\end{equation}
where $c=a+\delta a$ and the terms to first order in $\delta$ are
taken into account. In spheroidal coordinates, the only nonzero
component of $W_{\mu}$ in region $R_{N}$ is found to be (see
Ref.~\cite{Pudlak1})
\begin{equation}
W_{\phi}\approx g\cos\theta(1+\delta\sin^{2}\theta) +
G(1-\cos\theta)-\delta G\sin^{2}\theta
\cos\theta.\label{eq:19}\end{equation}
The external magnetic field $B$ is chosen to be pointed in the $x$ direction,
so that $\vec{A}=B\left(0,-z,y\right)/2$. One obtains
\begin{equation}
A_{\phi}=-\frac{1}{2}Bac\sin\theta\cos\theta\cos\phi,\label{eq:20}
\end{equation}
\begin{equation}
A_{\theta}=-\frac{1}{2}Bac\sin\phi.
\end{equation}
The Dirac matrices can be chosen to be the Pauli matrices,
$\gamma_1=-\sigma_2, \gamma_2=-\sigma_1$. By using the substitution
\begin{equation}
\left(%
\begin{array}{c}
  \psi_{A} \\
  \psi_B \\
\end{array}%
\right) =\sum_j \frac{e^{i(j+G)\phi}}{\sqrt{2\pi}}\left(%
\begin{array}{c}
  u_j(r) \\
  v_j(r) \\
\end{array}%
\right) ,j=0,\pm 1,\pm 2,\ldots
\end{equation}
we obtain the Dirac equation for functions $u_j$ and $v_j$ in the
form
$$
\left(-i\sigma_{1}\frac{1}{a}(\partial_{\theta}+\frac{\cot\theta}{2}+i\frac{1}{2}Ba^{2}\sin\phi)+\frac{\sigma_{2}}{a
\sin\theta}\left(j-m\cos\theta+\frac{1}{2}Ba^{2}\sin\theta\cos\theta\cos\phi\right)+\delta
\hat{\cal{D}}_{1}\right) \times
$$
\begin{equation}
\times\left(\begin{array}{c}
  u_j(\theta) \\
  v_j(\theta) \\
\end{array}
\right)=E\left(\begin{array}{c}
  u_j(\theta) \\
  v_j(\theta) \\
\end{array}
\right), \label{eq:21}
\end{equation}
where
\begin{equation}
\hat{\cal{D}}_{1}=-\frac{\gamma_{1}}{a}\sin\theta\left(j-2m\cos\theta\right)-\gamma_{1}\frac{B
a}{2}\sin^{2}\theta\cos\theta\cos\phi. \label{eq:22}
\end{equation}
The square of nonperturbative part of Dirac operator takes the form
\begin{eqnarray}
\hat{\cal{D}}_{0}^2&=&-\frac{1}{a^{2}}\left(\partial^{2}_{\theta}+\frac{\cos\theta}{\sin\theta}\partial_{\theta}-\frac{1}{4}-\frac{1}{4
\sin^{2}\theta}\right)+\frac{\left(j-m\cos\theta\right)^{2}}{a^{2}\sin^{2}\theta}\nonumber\\
&&+\sigma_{3}\frac{m-j\cos\theta}{a^{2}\sin^{2}\theta}+BV(\theta,\phi).
\label{eq:23}
\end{eqnarray}
Notice that in (11)-(13) the terms with $B^{2}$ and $\delta B$ were
neglected and
\begin{eqnarray}
V(\theta,\phi)&=&-i(\cos\phi\partial_\phi-\frac{\sin\phi}{2})\cot\theta-\sigma_3\frac{\cos\phi(1+\sin^{2}\theta)}{2\sin\theta}-i(\partial_\theta+\frac{1}{2}\cot\theta)\sin\phi \nonumber
\\
&&-\frac{m \cos^{2}\theta\cos\phi}{\sin\theta}.\label{eq:24}
\end{eqnarray}
Let us define
$V_{j,j'}^{m}=\langle\psi_{jn}^{m}|V(\theta,\phi)|\psi_{j'n}^{m}\rangle$,
where $\psi_{jn}^{m} $ are the eigenfunctions of
$\hat{\cal{D}}_{0} $. In the case of the first energy mode the
terms $V_{-1,-1}^{1/2}$, $V_{1,1}^{1/2}$, $V_{1,-1}^{1/2}$,
$V_{-1,1}^{1/2} $ are found to be zero and, therefore, the weak
magnetic field pointed in the x direction does not influence the
first energy level. For the second energy mode, only non-diagonal
terms $V_{-1,-2}^{1/2}$, $V_{1,2}^{1/2}$ differ from zero with
$V_{-1,-2}^{1/2}=-2.1$ and $V_{1,2}^{1/2}=2.1$. Notice that the following
wave functions were used for calculations of non-diagonal terms:
\begin{eqnarray}
\psi_{0,2}^{1/2} (z,\phi)= \frac{e^{i2\phi}}{\sqrt{2\pi}}\sqrt{\frac{32}{15}}\left(%
\begin{array}{c}
  -i\sqrt{\frac{2}{3}}\sqrt{1-z^{2}}(1+z)\\
   1-z^{2}\\
\end{array}%
\right) ,  \nonumber \\
\psi_{0,-2}^{1/2} (z,\phi)= \frac{e^{-i2\phi}}{\sqrt{2\pi}}\sqrt{\frac{32}{15}}\left(%
\begin{array}{c}
  i\sqrt{\frac{2}{3}}\sqrt{1-z^{2}}(1-z)\\
   1-z^{2}\\
\end{array}%
\right) ,
\end{eqnarray}

\begin{eqnarray}
\psi_{1,1}^{1/2} (z,\phi)= \frac{e^{i\phi}}{\sqrt{2\pi}}\sqrt{\frac{3}{5}}\left(%
\begin{array}{c}
  -i\sqrt{\frac{2}{3}}(2z^{2}+z-1)\\
   \frac{3}{2}z\sqrt{1-z^{2}}\\
\end{array}%
\right) ,  \nonumber\\
\psi_{1,-1}^{1/2} (z,\phi)= \frac{e^{-i\phi}}{\sqrt{2\pi}}\sqrt{\frac{3}{5}}\left(%
\begin{array}{c}
  i\sqrt{\frac{2}{3}}(-2z^{2}+z+1)\\
   \frac{3}{2}z\sqrt{1-z^{2}}\\
\end{array}%
\right) .
\end{eqnarray}
The low energy electronic spectrum of spheroidal fullerenes in this
case  takes the form
\begin{equation}
E_{jn}=E^0_{jn}+E^{\delta B_{x}},
\end{equation}
where
\begin{equation}
E^{\delta B_{x}}=\frac{\delta
\left(\hat{\Gamma}_{22}+\hat{\Gamma}_{11}\right)\pm\sqrt{\delta^{2}
\left(\hat{\Gamma}_{22}-\hat{\Gamma}_{11}\right)^{2}+4|Ba^{2}V_{1,2}|^2}}{4
E^0_{jn}} .\end{equation} Here $\hat{\Gamma}_{ii}$ are the diagonal
matrix elements (see Ref.~\cite{Pudlak2}). Table 1 shows
contributions to the first and second energy levels for YO-C$_{240}$
(YO means a structure given in \cite{yoshida,Lu}). As is seen, there
is a marked difference between the behavior of the first and second
energy levels in magnetic field. Indeed, in both cases the energy
levels become shifted due to spheroidal deformation. However, the
uniform magnetic field does not influence the first energy level.
The splitting takes place only for the second level. This is clearly
illustrated in Fig. 1 and Fig.2, which schematically show the
structure of the first and second levels in the uniform magnetic
field pointed in x direction. The case of the z-directed magnetic
field is also shown for comparison. We can conclude that there is a
possibility to change the structure of the electronic levels in
spheroidal fullerenes by altering the direction of magnetic field.
It would be interesting to test this prediction in experiment.

\section[]{The $SU(2)$ algebra}

To gain a better understanding of the difference between the spheroidal and spherical
cases, let us consider the angular-momentum operators for Dirac operator on the sphere
$S^2$ with charge $G$ and a total magnetic monopole $m$
\begin{equation}
\hat{L}_{z}=-i(\partial_{\phi}\mp i G),\label{eq:5.56}
\end{equation}
\begin{equation}
\hat{L}_{x}=i\sin\phi\partial_{\theta}+i\cos\phi
\frac{\cos\theta}{\sin\theta}\left(\partial_{\phi}\mp
iG-i\frac{m}{\cos\theta}\right)+\sigma_{z}\frac{\cos\phi}{2\sin\theta},\label{eq:5.57}
\end{equation}
\begin{equation}
\hat{L}_{y}=-i\cos\phi\partial_{\theta}+i\sin\phi
\frac{\cos\theta}{\sin\theta}\left(\partial_{\phi}\mp
iG-i\frac{m}{\cos\theta}\right)+\sigma_{z}\frac{\sin\phi}{2\sin\theta}.\label{eq:5.58}
\end{equation}
Here $-(+)$ signs correspond to the case of north (south) hemisphere, respectively. These
operators satisfy the standard commutations relations of the $SU(2)$
algebra
\begin{equation}
\varepsilon_{ijk}\hat{L}_{j}\hat{L}_{k}=i\hat{L}_{i}.\label{eq:5.59}
\end{equation}
For zero magnetic field, the square of the Dirac operator and $\hat{L}^{2}$
may be diagonalized simultaneously
\begin{equation}
\hat{\cal{D}}_{0}^2(B=0)=\hat{L}^{2}+\frac{1}{4}-m^{2}.\label{eq:5.60}
\end{equation}
For the magnetic field pointed in the z direction, the operator $V(\theta,\phi)$
can be expressed in the form
\begin{equation}
V(\theta,\phi)=\hat{L_{z}}+\left(\frac{\sigma_{3}}{2}-m\right)z,\label{eq:5.61}
\end{equation}
while in the case of the magnetic field pointed in the x direction one obtains
\begin{equation}
V(\theta,\phi)=-\hat{L_{x}}-\left(\frac{\sigma_{3}}{2}-m\right)x.\label{eq:5.62}
\end{equation}
Here $x$ and $z$ are the Cartesian coordinates:
$x=\sin\theta \cos\phi; \quad  z=\cos\theta$.
In this case, the square of the Dirac operator and operator $\hat{L}^2$ may also
be diagonalized simultaneously and their eigenvalues are interrelated
\begin{equation}
\langle E,n| \hat{L}^{2} |E,n \rangle
=l(l+1)=E^{2}-\frac{1}{4}=(n+|j|)(n+|j|+1),
\end{equation}
where
$l=n+|j|$.
The eigenstates of $\hat{\cal{D}}_{0}(B=0)$ have the form
$$
\Psi_{l,j}(x,\phi)=\frac{e^{i(j\pm G)\phi}}{\sqrt{2\pi}\Omega}
\sqrt{\frac{(l+j)!}{(l-j)!}}
$$
\begin{equation}
\times \left( \begin{array}{c}(1-x)^{- \frac{1}{2}(j-m-1/2)}(1+x)^{-
\frac{1}{2}(j+m+1/2)}\frac{d^{l-j}}{dx^{l-j}}(1-x)^{(l-m-1/2)}(1+x)^{(l+m+1/2)} \\
i sgn(E)(1-x)^{-\frac{1}{2}(j-m+1/2)}(1+x)^{-
\frac{1}{2}(j+m-1/2)}\frac{d^{l-j}}{dx^{l-j}}(1-x)^{(l-m+1/2)}(1+x)^{(l+m-1/2)}
\end{array} \right)
\end{equation}
where
$\Omega=2^{l}\sqrt{\Gamma(l-m+1/2)^{2}+\Gamma(l+m+1/2)^{2}}$ where
$\Gamma$ are the Gamma functions (see Ref.~\cite{Abrikosov}).
One can introduce the operators $\hat{L}_{-}$ and $\hat{L}_{+}$ so that
\begin{equation}
\hat{L}_{-}\Psi_{l,j}=\sqrt{(l+j)(l-j+1)}\Psi_{l,j-1},
\end{equation}
\begin{equation}
\hat{L}_{+}\Psi_{l,j}=\sqrt{(l+j+1)(l-j)}\Psi_{l,j+1},
\end{equation}
with
\begin{equation}
\hat{L}_{-}=-e^{-i\phi}\left(\partial_{\theta}-i\frac{\cos\theta}{\sin\theta}\left(\partial_{\phi}\mp
iG-i\frac{m}{\cos\theta}\right)-\frac{\sigma_{z}}{2\sin\theta}\right),
\end{equation}
\begin{equation}
\hat{L}_{+}=e^{i\phi}\left(\partial_{\theta}+i\frac{\cos\theta}{\sin\theta}\left(\partial_{\phi}\mp
iG-i\frac{m}{\cos\theta}\right)+\frac{\sigma_{z}}{2\sin\theta}\right).
\end{equation}
Let us transform these expressions to the Cartesian coordinates.
The corresponding transformation rules for spinors are~\cite{Abrikosov}
\begin{equation}
(\Psi)_{C}=V^{\dag} \Psi,
\end{equation}
and the Cartesian realization of operator $\hat{L}$ is
\begin{equation}
\hat{L}_{C}=V^{\dag} \hat{L}V,
\end{equation}
with
\begin{equation}
V=\exp\left(\frac{i\sigma_{y}}{2}\theta
\right)\exp\left(\frac{i\sigma_{z}}{2}\phi \right).
\end{equation}
For example, the Cartesian realization of $\hat{L}_{x}$
takes the form
\begin{equation}
(\hat{L}_{x})_{C}=i\sin\phi\partial_{\theta}+i\cos\phi
\frac{\cos\theta}{\sin\theta}\left(\partial_{\phi}\mp
iG-i\frac{m}{\cos\theta}\right)+\frac{\sigma_{x}}{2}.
\end{equation}
For $m=G=0$ it may be written in the well-known form
\begin{equation}
(\hat{L}_{x})_{C}=y \hat{p}_{z}-z \hat{p}_{y}+\frac{\sigma_{x}}{2}.
\end{equation}
\newpage

\section{Conclusion}

We have studied the influence of the uniform magnetic field on the
energy levels of spheroidal fullerenes. The case of the x-directed
magnetic field was considered and compared with the case
of the z-th direction. The z axis is defined
as the rotational axis of the spheroid with maximal symmetry.
The most important finding is that the splitting of the electronic
levels depends on the direction of the magnetic field.
Our consideration was based on the using of the eigenfunctions of the Dirac operator
on the spheroid, which are also the eigenfunction of $\hat{L}_{z}$.
Let us discuss this important point in more detail.

In the case of a sphere there is no preferable direction
in the absence of the magnetic field. The magnetic field sets a vector,
so that the z-axis can be oriented along the field. In this case, one has to use
such eigenfunctions of the Dirac operator which are also the eigenfunctions of $\hat{L}_{z}$.
For the x-directed magnetic field, the eigenfunction of both the Dirac operator
and $\hat{L}_{x}$ must be used. Evidently, the same results will be obtained in both cases.

The situation differs markedly for a spheroid. The spheroidal symmetry itself assumes the
preferential direction which can be chosen as the z-axis. In other words,
the external magnetic field does not define the preferable orientation.
The symmetry is already broken and, as a result, the case of the magnetic field
pointed in the $x$ direction differs from the case of the $z$-directed field.
For instance, there are no eigenfunctions which would be simultaneously the
eigenfunctions of both the Dirac operator on the spheroid and $\hat{L}_{x}$.
For this reason, the structure of the electronic levels is found
to crucially depend on the direction of the external magnetic field.

\vskip 0.2cm \vskip 0.2cm The work was supported in part by VEGA
grant 2/7056/27. of the Slovak Academy of Sciences, by the Science
and Technology Assistance Agency under contract No. APVT-51-027904
and by the Russian Foundation for Basic Research under Grant No.
05-02-17721.

\newpage

\begin{table}[htb]
\begin{center}
\begin{tabular}{l c c c c c }\hline
\bfseries\bfseries $YO-C_{240}$\quad \quad & $j$ & \bfseries
$E^{0}_{jn} (eV)$ & \bfseries $ E_{jn}^{\delta}(meV)(B=0)$
& \bfseries $E_{jn}^{\delta B_{x}}(meV)(B a^{2}=0.1)$\\
\hline\bfseries $n=0$, $m=1/2$
&1&1.094&10.5&10.5\\
&-1&1.094&10.5&10.5\\
\hline\bfseries $n=0$, $m=-5/2$
&3&1.89&3&3\\
&-3&1.89&3&3\\
\hline\bfseries  $n=0$, $m=1/2$
&2&1.89&28.4\\
&&&& 53/-16/\\
$n=1$ &1&1.89&8.8\\
\hline\bfseries  $n=1$, $m=1/2$
&-1&1.89&8.8\\
&&&& 53/-16/\\
$n=0$&-2&1.89&28.4\\
\hline
\end{tabular}
\end{center}
\caption{{\footnotesize The structure of the first and second energy
levels for YO-C$_{240}$ fullerene in uniform magnetic field. The
hopping integral and other parameters are taken to be $t=2.5\ eV$
and $V_F=3t\overline{b}/2\hbar$, $\overline{b}=1.45{\textmd{\AA}}$,
$\overline{R}=7.03{\textmd{\AA}}$, $SD=0.17{\textmd{\AA}}$,
$\delta=0.024$. $\overline{b}$ is average bond length,
$\overline{R}\ (\overline{R}=a)$ is average radius, $SD$ is standard
deviation from a perfect sphere (see Refs. [10,11]), so that
$\delta=SD/\overline{R}$.}} \label{tab}
\end{table}

\begin{figure}[htb]
  \centering
\includegraphics[width=0.37\textwidth]{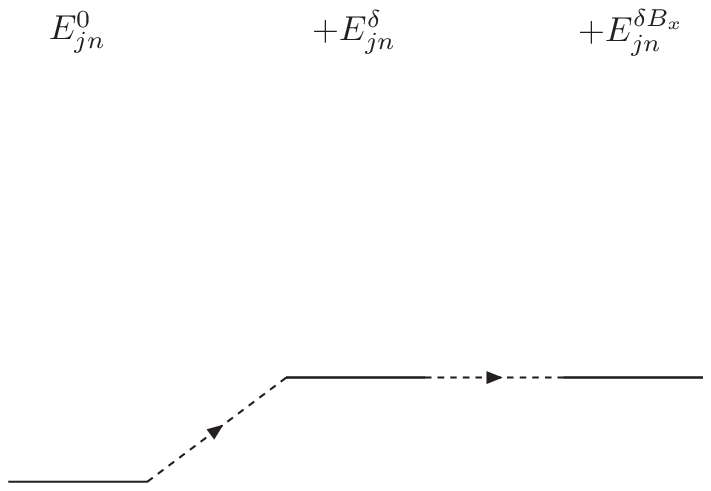}
  \quad\quad\quad\quad
  \includegraphics[width=0.30\textwidth]{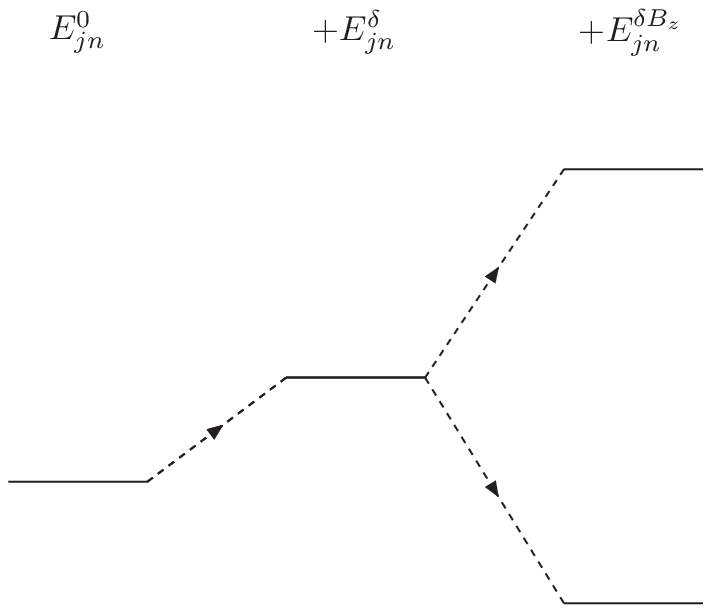}
\caption{ The schematic picture of the first electronic level $E_{jn}$
of spheroidal fullerenes in a weak uniform magnetic field pointed in
the x (left) and z (right) directions.}
\end{figure}

\begin{figure}[htb]
  \centering
  \includegraphics[width=0.40\textwidth]{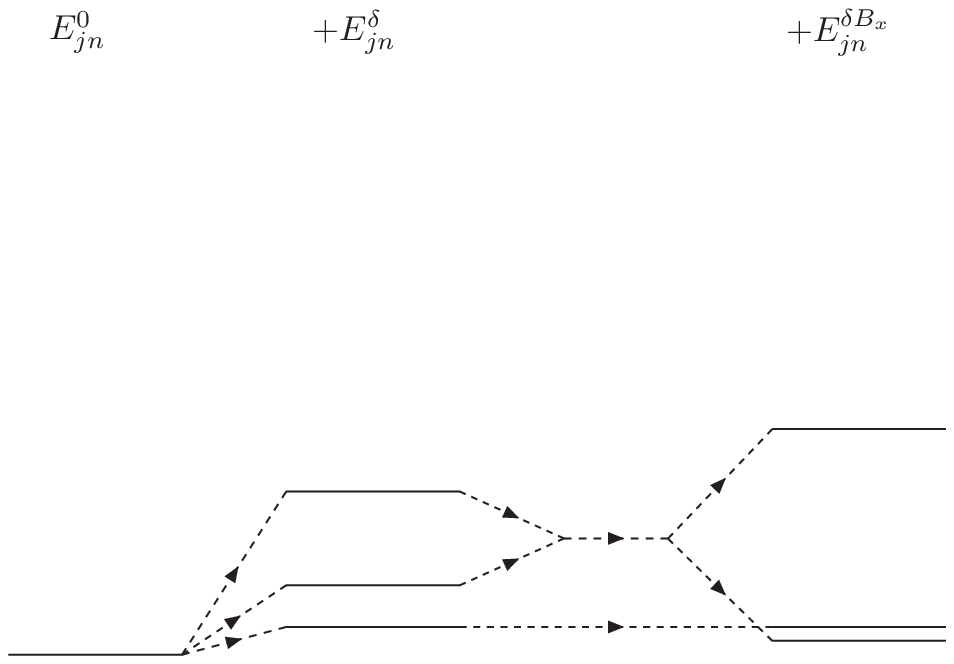}
  \quad\quad\quad\quad
  \includegraphics[width=0.28\textwidth]{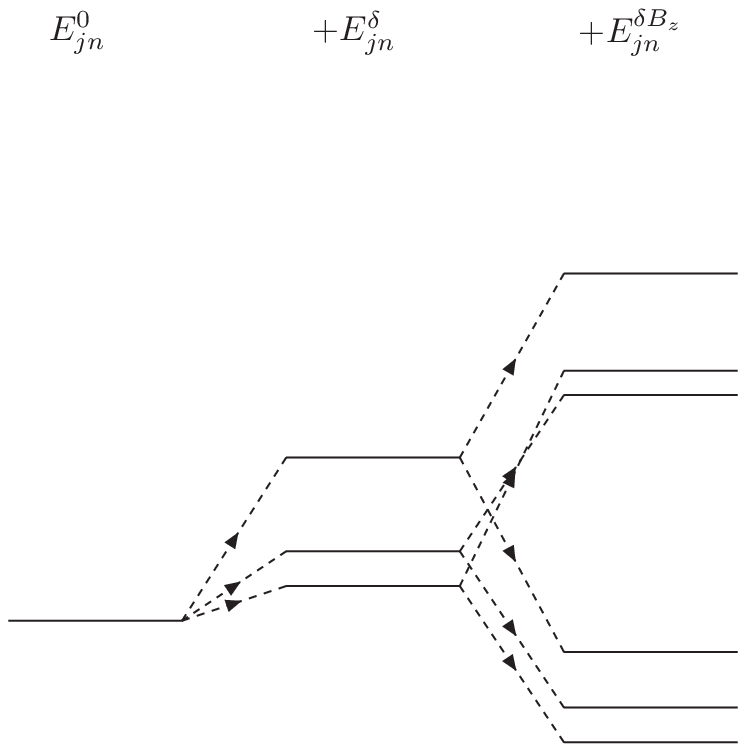}
  \caption{
The schematic picture of the second electronic level $E_{jn}$
of spheroidal fullerenes in a weak uniform magnetic field pointed in
the x (left) and z (right) directions.}
\end{figure}

\end{document}